 \documentclass[pra,amsmath,amssymb]{revtex4}

 \usepackage{graphicx}
 \usepackage{dcolumn}
 \usepackage{bm}
 \usepackage{subfigure}
 \newcommand{\YSO}{Y$_2$SiO$_5$}

 \newcommand{\ket}[1]{\left|#1\right\rangle} 
  
\begin{document}

 \title{Selecting ensembles for rare earth quantum computation}

 \author{J. J. Longdell}
 \email{jevon.longdell@anu.edu.au}
 \author{M. J. Sellars}
 \affiliation{Laser Physics Center, Research School of Physical
    Sciences and Engineering, Australian National University,
    Canberra, ACT 0200, Australia.}

  \date{August 29, 2002}
\begin{abstract}
  We discuss the issues surrounding the implementation of  quantum
  computation in rare-earth-ion doped solids. 
  We describe a practical scheme for two qubit gate operations which
  utilise experimentally available interactions between the qubits.
  Possibilities for a scalable quantum computer are discussed.
\end{abstract}

  \maketitle

\section{Introduction}

Quantum computing promises significant advantages over classical
computation for a number of important tasks such as factorising large
numbers \cite{shor94} and searching large databases \cite{grov96}.
Experimental quantum computation is difficult because of two competing
requirements. Firstly the qubits must be well isolated from
the environment to preserve coherence. Secondly they must have
interactions among themselves and with whatever is being used to
control and read out their states. In addition to this the power of a
quantum computer increases exponentially with its number of
qubits. For a quantum computer to outperform a
classical one it is more than likely a large number of qubits will be required.

There have been a number of demonstrations of quantum computation to date.
The physical systems used include liquid state NMR \cite{nmrcnottut}, atoms in
high-finesse cavities \cite{turc95} and ion traps \cite{cira95}. The
problem with all such schemes is that there has been trouble in
scaling the system to a large number of qubits. The handful of qubits
achieved to date are not sufficient for a technologically useful
device.

Great effort has been put into the nano-fabrication of devices that,
it is hoped, will enable scalable quantum computation. Recently two
interacting qubits have been demonstrated \cite{nakamura2} but the coherence
times in this system were short (approx 10 ns). 
Difficulties 
associated with these very short timescales provide a challenge for further scaling of the approach.

Rare-earth-ions provide another possible path to scalable quantum
computation, one where the coherence times are much longer and where
coherent effects have been studied for many years \cite{redbook}.
Decoherence effects, perhaps the most important issue in the operation of
a quantum computer are generally well understood in such systems.

One can summarise the attractive features of rare-earth-ion dopants for quantum
computing as below;

\begin{itemize}
\item Optical transitions with long coherence times, up to 2.6 ms,
  previously observed \cite{ultraslow}.
\item Long coherence times in their ground state nuclear spin
  transitions, up to 80 ms\cite{elliot}. 
\item Lifetimes for the nuclear spin states as long as
  several hours. \cite{yano91}
\item Strong coupling between the electronic states and nuclear spin
  states enabling the optical manipulation and readout of the spin
  states \cite{mlyn83,eric77}.
\item  Large inhomogeneous  broadening in
  the optical transitions, typically of the order of gigahertz enables many
  homogeneously broadened ensembles or even single ions to be addressed
  individually \cite{pryd00}.
\item Strong electric dipole-dipole interactions (as large as GHz)
  enabling coupling between the ions\cite{graf98}.
\item Unlike liquid state NMR, using optical pumping there are no
  problems in preparing the starting state.
\end{itemize}

In this paper we describe a practical method for achieving a two qubit device
using ensembles and discuss possibilities for scalability.

\section{Rare earth quantum computation}

The use of rare-earth-ions for quantum computation, and in particular
the implementation of two qubit gates using electric dipole-dipole
interactions, was first proposed by Sellars and
Manson\cite{sellarsandmansontalk}. These interactions are due to the fact
that the ground and optically excited states have different permanent
dipole moments. 

Following this a proposal for a complete architecture was given by
Ohlsson~et~al. \cite{ohls02}. The approach given here has the distinct
advantage that it is significantly less sensitive to inhomogeneous
broadening. This sensitivity to inhomogeneous broadening has been
recognised as the main weakness in attempts to improve the Ohlsson
scheme\cite{wese03}. 



\begin{figure}
  \centering
  \includegraphics[width=\textwidth]{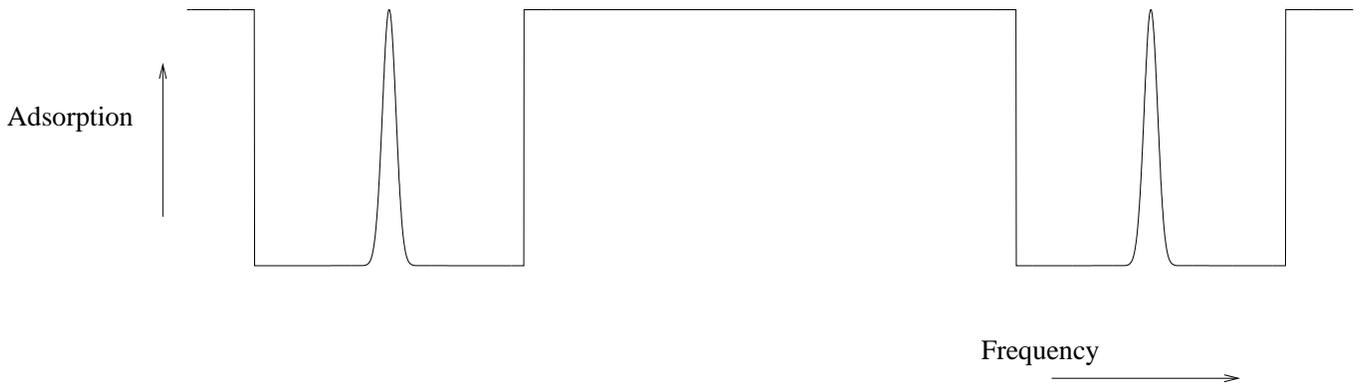}
  \caption{Sketch of absorbtion verses frequency for the two
  anti-holes used as interacting qubits in this scheme.}
  \label{fig:spike}
\end{figure}

Each qubit in such a scheme is a packet of ions chosen from the
inhomogeneous line based on their optical frequency, all ions that are
not part of this packet but have similar resonant frequencies would be
optically pumped to an auxiliary hyperfine level. Optical absorption
around such an anti-hole is shown schematically in
Fig.~\ref{fig:spike}. Ensembles of this type were first
realised, and their usefulness to quantum computing recognised 
by Pryde~et~al. \cite{pryd00}. A clear demonstration that such
systems perform well as qubits is given in \cite{long02}.

There is a strong analogy between the ensemble quantum computation using
rare earth ions and liquid state NMR. In both cases the measurements
on ensembles can be made by monitoring the coherent emission. In both
cases qubits can be addressed based on their resonant frequency and
the interaction between the qubits is weak compared to available
driving Rabi frequencies. It should be noted that in the case of rare
earth ions this weak interaction is a result of using ensembles and
the interaction between the ions can be large.

As already stated, due to optical pumping, rare earth ion quantum
computation doesn't share the initialisation problem of NMR. Further
to this the rare earth ions have optically resolvable hyperfine
structure allowing the transfer of quantum information from the
optical transitions to ground state hyperfine transitions. When the
quantum information is stored in this ground state structure it can be
stored for long periods of time (80 ms\cite{elliot}) and it is
insensitive to the electric dipole-dipole interaction.

The challenge for rare earth quantum computing is the inhomogeneity in
the strengths of the interactions between qubits. This is not a problem
in liquid state NMR, since the nuclear spins in a molecule have fixed
relative positions whereas the rare earth dopant ions in an ordinary
crystal are randomly distributed.

\begin{figure}
  \centering
  \includegraphics{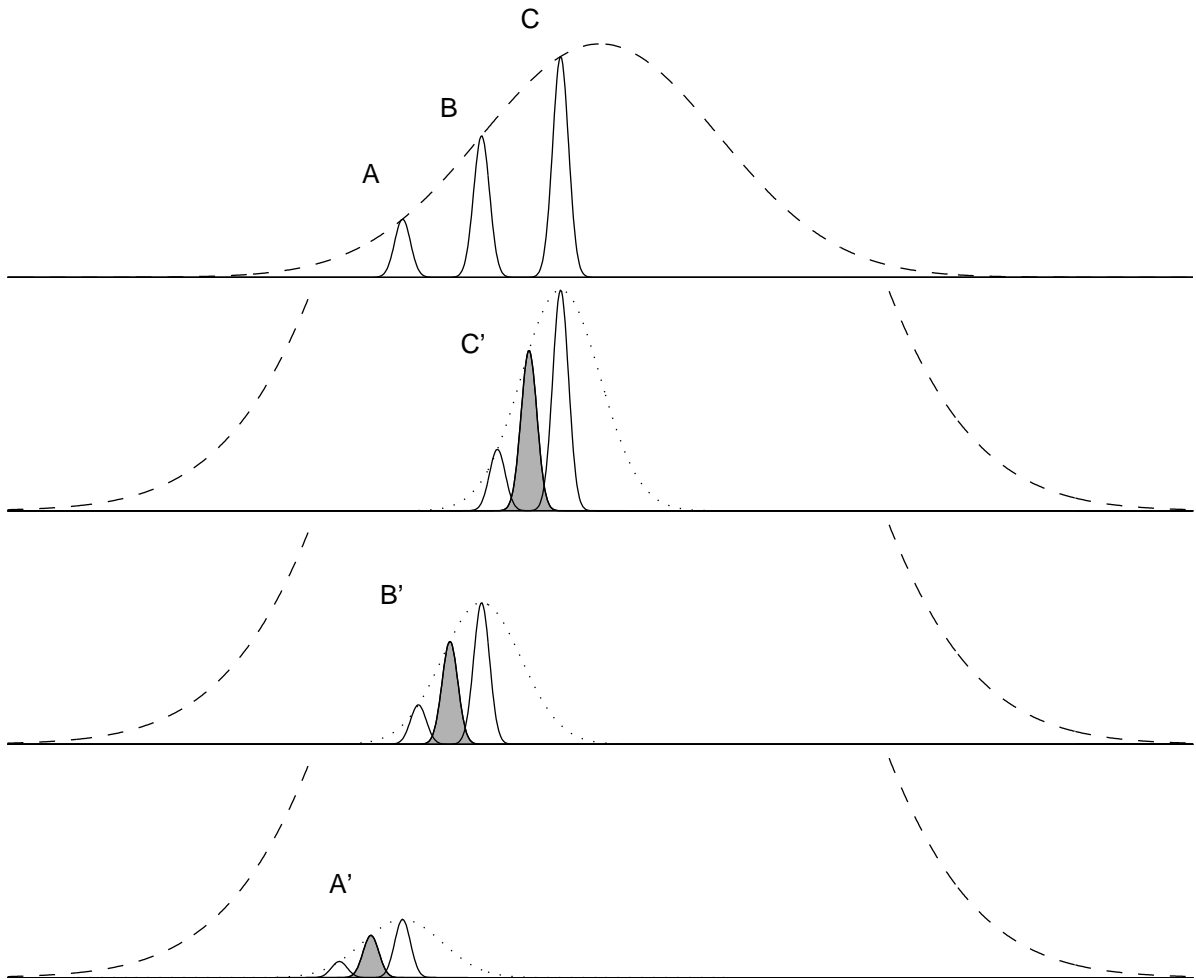}
  \caption{A diagram showing the effect on the frequencies of 
  of the ions of an anti-hole upon the excitation of another. Before
  the other anti-hole is excited (top) the inhomogeneously broadened
  feature is made up of a whole lot of homogeneously broadened
  features (A, B and C). The excitation of another group of ions will
  cause random shifts in the resonance frequencies of the ions leading
  to broadening of these ensembles. These are shown in the bottom
  three plots. The technique described here selects out the shaded
  areas and uses the resulting deterministic frequency shift for 
  quantum computing operations.} 
  \label{fig:neilfig}
\end{figure}

The effect on the frequencies of the ions of one anti-hole upon the
excitation of another is shown in Fig.~\ref{fig:neilfig}. The
homogeneous linewidths of rare earth systems can be as small as 100
Hz\cite{ultraslow}. Individual homogeneous packets are represented by
A, B and C in the top trace.  The excitation of a second anti-hole
causes a broadening in what would have been homogeneously broadened
packets denoted by A', B' and C'. How much the frequency of an ion
shifts depends on how close it is to an ion in the perturbing
anti-hole.  Obviously the number density of perturbing ions increases
with the spectral width of the perturbing anti-hole.  Thus the
broadening caused by the interaction increases also with the spectral
width of the perturbing anti-hole. A simple analysis \cite{nils02}
shows that this broadening is {\em proportional} to the spectral width
of the perturbing ions and approximately 20 times smaller. This agrees
with measurements made of this interaction \cite{long02}.


Coherent transient techniques \cite{redbook} have been used
extensively in rare earth systems to probe the rare-earth-ion system 
with greater resolution than would be allowed by the
inhomogeneous broadening using conventional techniques. Here we
provide a method for achieving both the ion selection and 
computation that is based on photon echo sequences. 
The strength of this technique is that it allows the use of
interaction strengths smaller than the inhomogeneous linewidth of
anti-holes that are used as qubits. As mentioned above the mean
interaction strength between the ions of two anti-holes is a small
fraction of their homogeneous width.  Earlier proposals \cite{ohls02}
are limited to frequency
shifts bigger than the inhomogeneous linewidth and are thus forced to work
with a very small subset of the
ions. A level of 0.3\% has been estimated \cite{nils02} by one of the
proponents of the scheme. The requirement this puts on the level
background ions is a problem in practice. 

\begin{figure}
\includegraphics{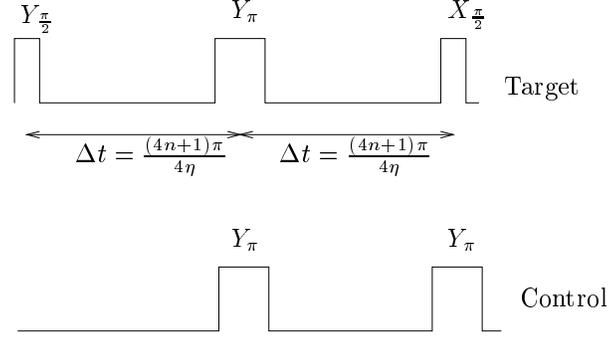}
\caption{\label{fig:pulseseq}Pulse sequence to achieve a CNOT gate in
  the presence of inhomogeneous broadening. $X_{\pi}$ is a pulse that
  causes a $\pi$ rotation about the $x$ axis on the Bloch sphere.}
\end{figure}

The pulse sequence to achieve a CNOT operation between the two ions is
illustrated in figure \ref{fig:pulseseq}, and can be understood with
the help of the following Hamiltonian.
In the appropriate interaction picture the Hamiltonian for two ions in
two separate anti-holes is give by.
\begin{equation}
  \label{eq:Ham3}
  H = \frac{\delta_1}{2} Z_1 + \frac{\delta_2}{2} Z_2 + \frac{\eta}{2} Z_1 Z_2
\end{equation}
Here $\delta_i$ represents the detuning from the center anti-hole and $Z_i$
is the Pauli-Z operator ($Z = [{1}\ {0};\ {0}\ {-1}]$) for the ion
$i$.  The strength of the interaction between the two ions is
described by $\eta$.

 The first pulse puts the
target ion on the equator of the Bloch sphere where it then precesses around
the equator at a rate given by its detuning from resonance. This
detuning is given by the sum of $\delta_1$ and the effect of the
interaction. Without the $\pi$ pulse on the control qubit at the
middle of the gate operation a $\pi$ pulse to the target gate would
refocus this precession. This would lead to the target ion being at
the same position on the equator of the Bloch sphere as when it
started. The application of the $\pi$ pulse to the control ions stops
the re-phasing of the precession due to the interaction. If the
waiting time is equal to an odd multiple of $\frac{\pi}{4\eta}$ then
the two trajectories on the block sphere corresponding to the control
ion being initially in the states $\ket{0}$ and $\ket{1}$ end up
separated by angle $\pi$. They can then easily be mapped to the ground
and excited states with the appropriate $\pi/2$ pulse.  Such a pulse
sequence would be useful in liquid state NMR if there were
limitations imposed  by
the inhomogeneity of the DC magnetic field.

The unitary evolution operator describing the gate's operation can be
expressed as the product of seven elementary operations.
\begin{equation}
  \label{eq:Udefinition}
  U = U_7U_6U_5U_4U_3U_2U_1
\end{equation}
where
\begin{eqnarray}
  \label{eq:Usdefinition}
  U_1 &=& \exp\left(\frac{-iY_1\frac{\pi}{2}}{2}\right)\\
  U_2 &=& \exp\left(-iH \Delta t\right)\\
  U_3 &=& \exp\left(\frac{-iY_1\pi}{2}\right)\\
  U_4 &=& \exp\left(\frac{-iY_2\pi}{2}\right)\\
  U_5 &=& \exp\left(-iH \Delta t\right)\\
  U_6 &=& \exp\left(\frac{-iX_1\frac{\pi}{2}}{2}\right)\\
  U_7 &=& \exp\left(\frac{-iY_2\pi}{2}\right)
\end{eqnarray}
this gives us, independent of $\delta_{1,2}$
\begin{equation}
(-1)^{\frac{1}{4}}
  \begin{bmatrix}
\sin(\theta) & -i\cos(\theta) & 0 & 0 \\
-\cos(\theta) & i\cos(\theta) & 0 & 0 \\
0 & 0 & \cos(\theta) & -i\sin(\theta) \\
0 & 0 & -\sin(\theta) & -i\cos(\theta)\\
\end{bmatrix}
\end{equation}
where $\theta = 2\eta\Delta t$, for $\Delta t = \frac{(4n+1) \pi}{4}$
this becomes 
 \begin{equation}
   (-1)^{\left(\frac{1}{4}+n\right)}
    \begin{bmatrix}
   -1 & 0 & 0 & 0 \\
    0 & i & 0 & 0 \\
    0 & 0 & 0 & i \\
    0 & 0 & 1 & 0
 \end{bmatrix}
 \end{equation}

The difference between this and the evolution matrix describing the
CNOT operation
\begin{equation}
    \begin{bmatrix}
    1 & 0 & 0 & 0 \\
    0 & 1 & 0 & 0 \\
    0 & 0 & 0 & 1 \\
    0 & 0 & 1 & 0
 \end{bmatrix}
\end{equation}
is simply rotations about the $Z$ axes of the Bloch spheres of the
qubits. These need not be physical operations but instead changes in
the definition of the phases for the qubits \cite{nmrcnottut}.

In order to create an ensemble with interaction strengths given by
$\eta \Delta t = (4n+1)\pi/4$ the gate operation can be applied
repeatedly with both ensembles initially in their ground states with a
pause between them of the order of the spontaneous emission time to allow
the system to relax. The target ions that see the correct interaction
strengths will still be in the ground state at the end of the
operation while the others will have some population in the excited
state. Repeated application of such sequences swapping the roles of
target and control qubits will optically pump all the ions
that don't see the correct interaction strength into an auxiliary
hyperfine level in a way analogous to spectral holeburning.

\begin{figure}
    \subfigure[]{\includegraphics[width=0.148\textwidth]{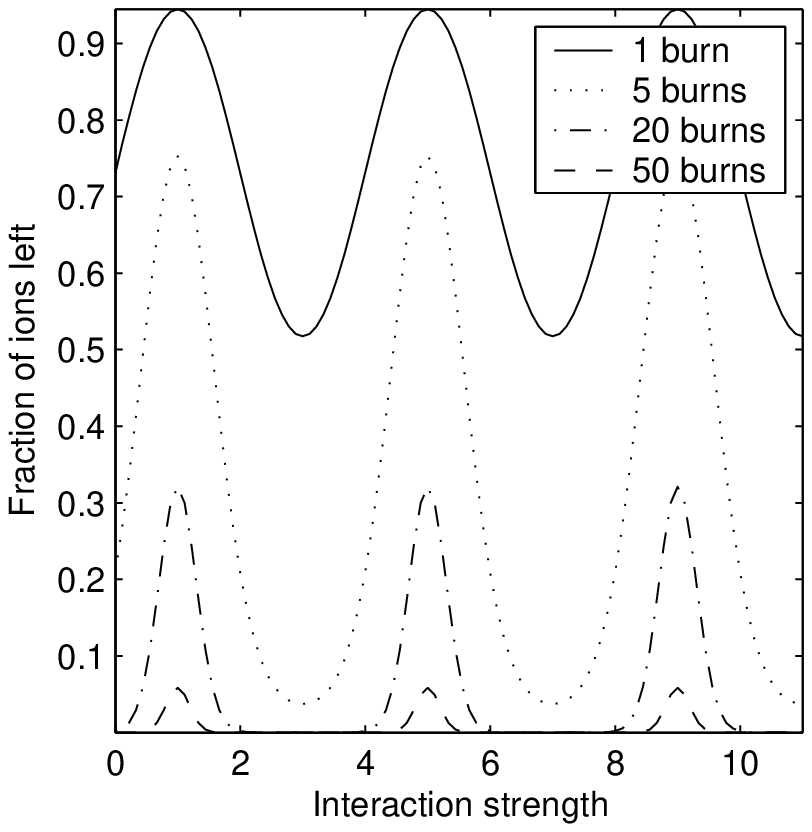}\label{fig:settinga}}
    \subfigure[]{\includegraphics[width=0.148\textwidth]{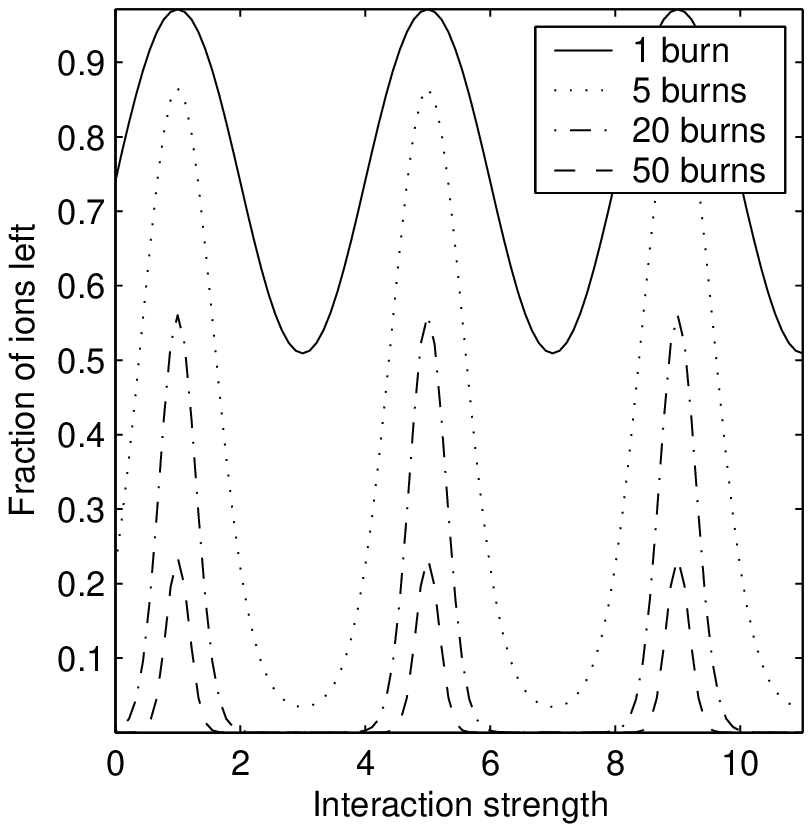}\label{fig:settingb}}
    \subfigure[]{\includegraphics[width=0.148\textwidth]{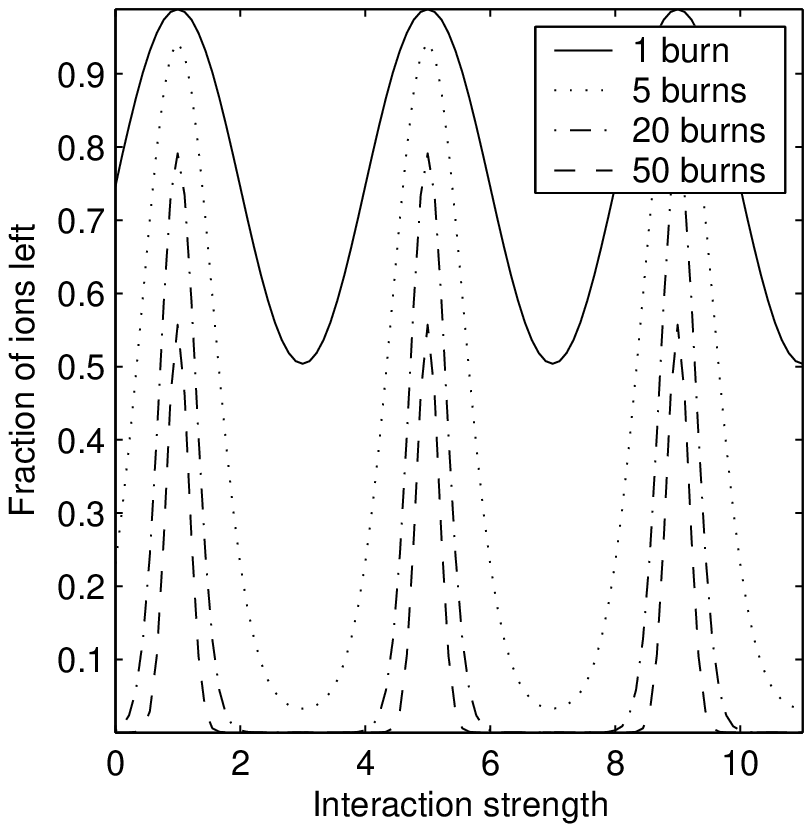}\label{fig:settingc}}
    \subfigure[]{\includegraphics[width=0.148\textwidth]{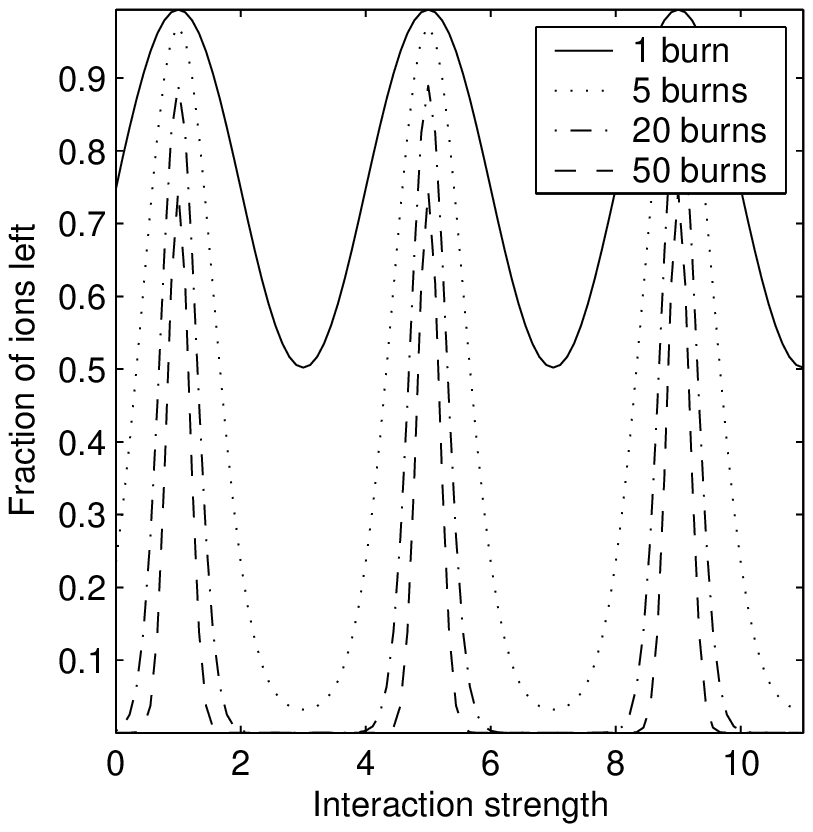}\label{fig:settingd}}
  \caption{Selecting target ions with the desired interaction strength
    using repeated application of the CNOT gate. Each graph shows the
    fraction of ions left at a particular interaction strength after
    1, 5, 20 and 50 gate operations. The time for free evolution
    between pulses in the CNOT gate was $\Delta t = \pi/4$. The effect
    of spontaneous emission was included with the spontaneous emission
    rates being: (a) $\gamma = \frac{1}{10}$, (b) $\gamma =
    \frac{1}{20}$, (c) $\gamma = \frac{1}{50}$, (d) $\gamma =
    \frac{1}{100}$.}
  \label{fig:interaction_strength_burning}
\end{figure}

Results of modeling of the interaction-strength-holeburning process
are shown in figure \ref{fig:interaction_strength_burning}. The
branching ratio for an atom to spontaneously emit back down into the
state $\ket{0}$ versus another hyperfine level was taken to be one
half.

One problem in this method is achieving high fidelity
gate operation in the presence of what is still reasonably large
inhomogeneity in the interaction strength. This problem can be
overcome using the results of Jones \cite{jone03}.
Jones pointed out that the problem of inhomogeneity in interaction
strength for two qubit gates is exactly analogous to the problem of
inhomogeneity in Rabi frequency for single qubit operations. This
means that methods analogous to the ``composite pulses'' used to
overcome these problems can be used. In the absence of other errors,
using such techniques would enable a fidelity better than $10^{-6}$
in the presence of 10\% inhomogeneity in the interaction strengths.

This leads us to the conditions that must be satisfied in order to
demonstrate the CNOT using this refocusing process. Firstly, the
average interaction between the ions must be at least an order of
magnitude larger than the spontaneous emission rate. This
condition is more easily satisfied when the starting ensembles are
wider anti-holes. The second condition is that the spectral width of the
ensembles must be small enough compared to the available laser power
that the pulses applied have the same effect to the whole ensemble.
For the example of Eu:\YSO\ if we burn anti-holes of the order of 100kHz
wide then we expect to have appreciable numbers of ions with
interaction strengths of the order of 10kHz. This is large compared to
the $\approx500$ Hz spontaneous emission rate. With 500 mW of laser
power focused with a 10cm lens, Rabi frequencies of the order 1 MHz
can be easily achieved.

A second condition is that the laser used to produce the pulses is
phase stable for the time of the experiment ($\approx 200\mu s$).
While making a laser this stable is non-trivial, a laser with
sufficient stability designed specifically for the spectroscopy of
rare earth ion systems exists \cite{pryd00,long02,dyke99}. 
In conclusion, this suggests that the demonstration of such a
two qubit gate is achievable with current technology. 

\section{\label{sec:inhomogeniety} The problem of inhomogeneity in
  interaction strength}

In a sense the problem of inhomogeneity is the only problem to
overcome in rare earth quantum computation. The ions have
experimentally verified long coherence times and large interaction
strengths. Further to this, for ensembles, measurements akin to those
used in liquid state NMR are available.  In the approach suggested
above ions are selected from a macroscopic collection based on their
resonant frequency and interaction strengths. The criteria for which
ion groups are acceptable to be part of the ensemble of quantum
computers gets increasingly more stringent with the number of qubits.
As mentioned above this will lead to exponentially fewer ions fitting
the criteria as the number of qubits increases.

In order to make rare-earth quantum computing scalable there is a need
for a better way of overcoming this inhomogeneity. The 
possibilities discussed here are the use of single ion spectroscopy either
directly or by coupling to other ions, and the use of ``solid
state molecules''.

In liquid state NMR based quantum computing you are dealing with an
ensemble of ``computers'' just as for the rare earth scheme described
above. The problems of inhomogeneity are avoided because each computer is
a molecule which is identical to all the others. It should be
emphasised that rare-earth quantum computing does not share the
initialisation problem which causes NMR quantum computing to become
untenable for large numbers of qubits. A material can be imagined
where there are large numbers of identical collections of rare earth
dopants.  All the rare earths are strongly electro-positive with their
bonding to other atoms essentially ionic in nature \cite{cotton} and
as such incorporating them in a molecular solid would be difficult.
One possibility for realising something akin to ``solid state
molecules'' is adding defects to a stoichiometric sample. Crystals
containing stoichiometric amounts of europium can still exhibit
relatively long coherence times so long as the inter-europium space is
large within the crystal \cite{eu_tungstate}.  The europiums close to
a defect would provide one member of an ensemble of an identical groups
of europium ions. The defect would shift the resonant frequencies of the
members of this group to differing amounts allowing them to be
manipulated based on their optical frequencies. The fact that they are
shifted out of resonance with the bulk europiums ions should also increase
their coherence times.

\begin{figure}
  \centering
  \includegraphics[height=0.3\textheight]{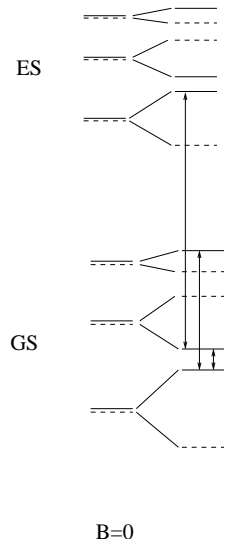}
  \caption{Energy levels diagram showing cyclic transition. ES denotes
  the optically excited state and GS denotes the ground state.} 
 \label{fig:cyclic}
\end{figure}

Another way of overcoming the inhomogeneity in interaction strength is by
abandoning the use of ensembles. In such a situation the computer
would consist of a cluster of ions. Because the ions are selected
based on their frequency rather than their precise positions and the
gate operations can be tailored to given interaction strengths, no
complex fabrication would be required. This leaves the problem of
detection of single ions.  Spectroscopic measurements of single NV
centres in diamond has been demonstrated \cite{grub97}. While the
lower oscillator strengths for rare earth ions would push current
detector technology, it may be possible to detect the 1000 photons/sec
produced when strongly driving an optical transition. How long this
emission lasts depends on the rate at which the population gets optically pumped
into other hyperfine levels. For free atoms/ions strong selection
rules result in cyclic transitions where the optically excited state
only decays into the ground state from which it is being driven. These
cyclic transitions are harder to come across in the solid
state but one option is using a site with a symmetry axis (for example
LaCl$_3$). At zero magnetic field the hyperfine structure is
described by a  psuedo-quadrupole Hamiltonian of the form
\begin{equation}
  \label{eq:pseudoquad}
  H = D(I_z^2 - I^2/3) + E(I_x^2-I_y^2)
\end{equation}
For reasons of symmetry the $z$ axes are the same for each electronic
state of the crystal. The eigenvalues for the above Hamiltonian can be
broken into two groups one of which consists of linear combinations of
the $I_z$ states $\{\ket{1/2},\ket{-3/2},\ket{5/2}\}$ and the other
linear combinations of the states
$\{\ket{-1/2},\ket{3/2},\ket{-5/2}\}$. These two groups are closed
under the operations of driving and spontaneous emission leading to a
cyclic transition that can be used to read out the state if RF
repumping fields are applied that connect all the other members of the
group. One complication is that the hyperfine states are two fold
degenerate at zero field with each pair consisting of a member of each
group. This degeneracy can be lifted without affecting the closed
nature of the groups by applying a magnetic field along the $z$
direction. This is shown on an energy level diagram in
Fig.~\ref{fig:cyclic}.

\section{Conclusion}

We have described a practical method for two-qubit quantum computing
demonstrations using ensembles. However, we conclude that such an
approach will not be scalable to a large number of qubits. Various
possibilities for scaling were discussed.


\end{document}